\documentclass[twocolumn,aps,prl,preprintnumbers]{revtex4}
\usepackage{}
\usepackage{float}
\usepackage{amsfonts}
\usepackage{bbm}
\usepackage[latin9]{inputenc}
\usepackage{amsmath}
\usepackage{amssymb}
\usepackage{graphicx}
\usepackage{mathrsfs}
\usepackage{amsfonts}
\usepackage{amsthm}
\usepackage{color}
\usepackage{txfonts}
\usepackage[colorlinks=true,citecolor=blue,linkcolor=blue,urlcolor=blue,anchorcolor=blue]{hyperref}%
\hypersetup{colorlinks=true,citecolor=blue,linkcolor=blue,urlcolor=blue}
\newcommand{\sign}[1]{\mathrm{sgn}(#1)}
\providecommand{\U}[1]{\protect\rule{.1in}{.1in}}
\makeatletter
\@ifundefined{textcolor}{}
{
\definecolor{BLACK}{gray}{0}
\definecolor{WHITE}{gray}{1}
\definecolor{RED}{rgb}{1,0,0}
\definecolor{GREEN}{rgb}{0,1,0}
\definecolor{BLUE}{rgb}{0,0,1}
\definecolor{CYAN}{cmyk}{1,0,0,0}
\definecolor{MAGENTA}{cmyk}{0,1,0,0}
\definecolor{YELLOW}{cmyk}{0,0,1,0}
}
\makeatother
\begin{document}
\title{Nonreciprocal spin waves driven by left-hand microwaves}
\author{Zhizhi Zhang}
\author{Zhenyu Wang}
\author{Huanhuan Yang}
\author{Z.-X. Li}
\author{Yunshan Cao}
\author{Peng Yan}
\email[Corresponding author: ]{yan@uestc.edu.cn}
\affiliation{School of Electronic Science and Engineering and State Key Laboratory of Electronic Thin Films and Integrated Devices, University of Electronic Science and Technology of China, Chengdu 610054, China}

\begin{abstract}
It is a conventional wisdom that a left-hand microwave cannot efficiently excite the spin wave (SW) in ferromagnets, due to the constraint of angular momentum conservation. In this work, we show that the left-hand microwave can drive nonreciprocal SWs in the presence of a strong ellipticity-mismatch between the microwave and precessing magnetization. A compensation frequency is predicted, at which the left-hand microwave cannot excite SWs. Away from it the SW amplitude sensitively depends on the ellipticity of left-hand microwaves, in sharp contrast to the case driven by right-hand ones. By tuning the microwave frequency, we observe a switchable SW non-reciprocity in a ferromagnetic single layer. A mode-dependent mutual demagnetizing factor is proposed to explain this finding. Our work advances the understanding of the photon-magnon conversion, and paves the way to designing diode-like functionalities in nano-scaled magnonics.
\end{abstract}

\maketitle
\section{INTRODUCTION}
Magnonics is an emerging field aiming for the future low-loss wave-based computation \cite{Barman2021,Mahmoud2020,Demidov2017,Grundler2016,Chumak2015,Lenk_2010,Serga2010}. Among the splendid magnonic functionalities, chirality and non-reciprocity serve as the basic building blocks \cite{Chen2021,Szulc2020,Grassi2020,Lan2015,Jamali2013} for the integrated magnonic circuits since the spin precession is innately chiral \cite{Pirro2021,Kruglyak2021,Kruglyak_2010}. The non-reciprocity can root in the magneto-dipolar interaction via, for example, the well known Damon-Eshbach (DE) geometry \cite{Damon1961,Camley1987,An2013,Kwon2016}, bilayer magnet and inhomogeneous thin film \cite{Ishibashi2020,GallardoPRAppl2019,AnPRAppl2019,Gladii2016,GallardoNJP2019,Borys2021,MacedoAEM2021,SadovnikovPRB2019}, and magnetic heterostructure in the presence of magneto-elastic or magneto-optic coupling \cite{TatenoPRAppl2020,Shah2020,ZhangPRAppl2020,WangPRL2019}. However, with the isotropic exchange interaction dominating in the microscale region \cite{WangNE2020,Chumak2014,Mohseni2019}, the dipolar effects, followed by the induced non-reciprocity, are vanishingly small \cite{Wong2014}. The non-reciprocity can also emerge in the chiral edge states of elaborately devised topological magnetic materials or spin-texture arrays, which are robust to defects and disorders \cite{WangPRAppl2018,WangPRB2017,LiPR2021,LiPRB2018,MookPRB2014,Shindou2013}. But it requires specific lattice designs and complicated couplings between atoms or elements, and the confined magnon (the quantum of spin wave) channels at the edges reduce the usage of the magnetic systems. Another origin for the non-reciprocity comes from the Dzyaloshinskii-Moriya interaction (DMI) \cite{Udvardi2009}. Yet, the effect is negligibly weak in ferromagnetic insulators, like yttrium iron garnet (YIG, Y$_3$Fe$_5$O$_{12}$) \cite{WangPRL2020}. Additional heavy metal structures can introduce a sizable DMI \cite{Bouloussa2020,Hrabec2020,GallardoPRL2019}, but inevitably bring remarkably increased damping and Joule heating \cite{SunPRL2013}.

To realize an efficient excitation of the non-reciprocal short-wavelength dipolar-exchange or even pure exchange spin waves (SWs) in ferromagnetic insulators for miniaturizing magnonic devices, several promising methods have been suggested \cite{AuAPL2012Res,AuAPL2012Nano,ChenPRB2019,ChenACS2020,WangNR2020,Sushruth2020, FrippPRB2021}. Conventionally, the coherent SW excitation harnesses the microwave antennas with the exciting field linearly polarized and uniform across the film thickness. Since the in-plane component of microwave fields dominantly contributes to the excitations, it is solely accounted in the analysis \cite{Dmitriev1988,Schneider2008,Demidov2009,Kasahara2017}. By contrast, the dynamic fields generated by micro-magnetic structures are not only highly localized at interfaces favoring the short-wavelength SWs excitation \cite{Yu2016,Liu2018,Che2020}, but also polarized with complex chiralities. Yu {\it et al.} have reported an analysis for the chiral pumping (excitation) of exchange magnons in YIG into (from) the proximate magnetic wires via directional dipolar interactions \cite{YuPRB2019,YuPRL2019}. A selection rule is adopted that circular magnons and photons with the same (opposite) chiralities are allowed (forbidden) to interact \cite{ZhangPRAppl2020}. One critical issue noteworthily rises how the microwave fields with contrary chirality excite the propagating SWs.

In this work, we theoretically investigate the propagating SWs in ferromagnetic films excited by microwave fields with generic chiralities. We find that the left-hand microwave can drive SWs because of the ellipticity mismatch between microwave and dynamic magnetization, which extrapolates the aforementioned selection rule for the magnon-photon conversion. Since the contributions of the in-plane and out-of-plane components of left- (right-) hand microwave fields are destructive (constructive) superposed, we introduce an analog to the common and differential signals in the differential amplifier. Surprisingly, we find a compensation frequency where no SWs can be excited by left-hand microwaves with certain ellipticity. We propose a proof-of-concept strategy for generating non-reciprocal SWs via applying the left-hand local microwave unevenly across the film thickness. A directional mutual demagnetizing factor is suggested to understand the emerging switchable SW chirality that depends on the microwave frequency. This proposal makes full use of the magnetic structures without breaking the symmetry of the dispersion relations and increasing the damping, which is superior to other methods. Our work lays a foundation of employing the chiral excitation for magnonic diodes in nano scales.

The paper is organized as follows. In Sec. \ref{Chiral}, we present the characteristics of the chiral excitation of SWs via a combination of theoretical analysis and numerical simulations. The strategy for nonreciprocal SWs excitations is proposed and demonstrated in Sec. \ref{Nonreciprocal}. Discussions and conclusions are drawn in Sec. \ref{Conclusion}.
\section{CHARACTERISTICS OF SPIN WAVES DRIVEN BY CHIRAL EXCITATIONS}\label{Chiral}
\subsection{MODELLING AND DISPERSION RELATION}\label{MODEL}
\begin{figure}
  \centering
  \includegraphics[width=0.40\textwidth]{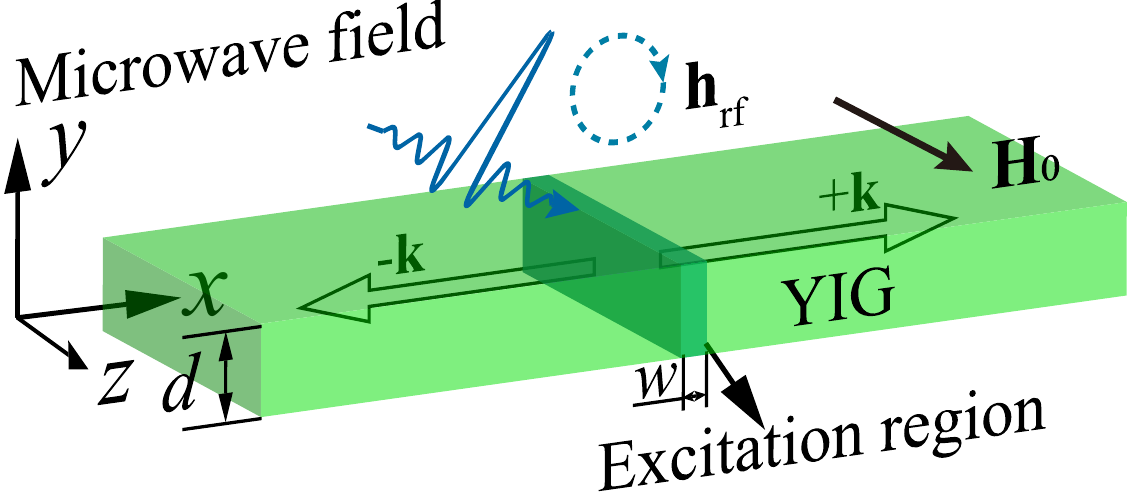}\\
  \caption{Schematic of the chiral excitation of SWs. The chiral microwave field ${\bf h}_{\rm rf}$ is locally applied in the patched green region. The SWs are propagating along $x$ direction indicated by the hollow arrows.} \label{fig0}
\end{figure}

We consider a ferromagnetic layer YIG with thickness $d$ extended in the $x-z$ plane and magnetized along $z$ direction by the bias magnetic field ${\bf H}_0={H}_0 {\bf z}$ (see Fig. \ref{fig0}). The microwave field ${\bf h}_{\rm rf}$ for the SWs excitation is centered at $x=0$ and located in the region with width $w$. {\color{red}The characteristics of SWs propagating along $x$ direction, i.e., the DE geometry \cite{Damon1961}, are investigated.} In the calculations, we set $d =$ 40 nm, $H_0 =$ 52 mT, and $w =$ 10 nm if not stated otherwise. The narrow excitation width ensures ${\bf h}_{\rm rf}$ comprises multiple wave vector in a wide range within $2\pi/w$ \cite{Sushruth2020,Stancil_2009}. Micromagnetic simulations are performed using MuMax3 \cite{Vansteenkiste2014} to verify the derived theories. {\color{red}The systems are meshed by cells with dimensions equal to $2\times2\times100~{\rm nm^3}$. Periodic boundary conditions ($\rm {PBC}\times200$) in the $z$ direction are applied, which means that the film is practically infinite in the $z$ direction. Absorbing boundary conditions are applied by adding the attenuating areas (not shown in the figure) where the $\alpha$ gradually increases to 0.25 to avoid the reflection at the two ends of simulated systems. }

\begin{figure}
  \centering
  \includegraphics[width=0.40\textwidth]{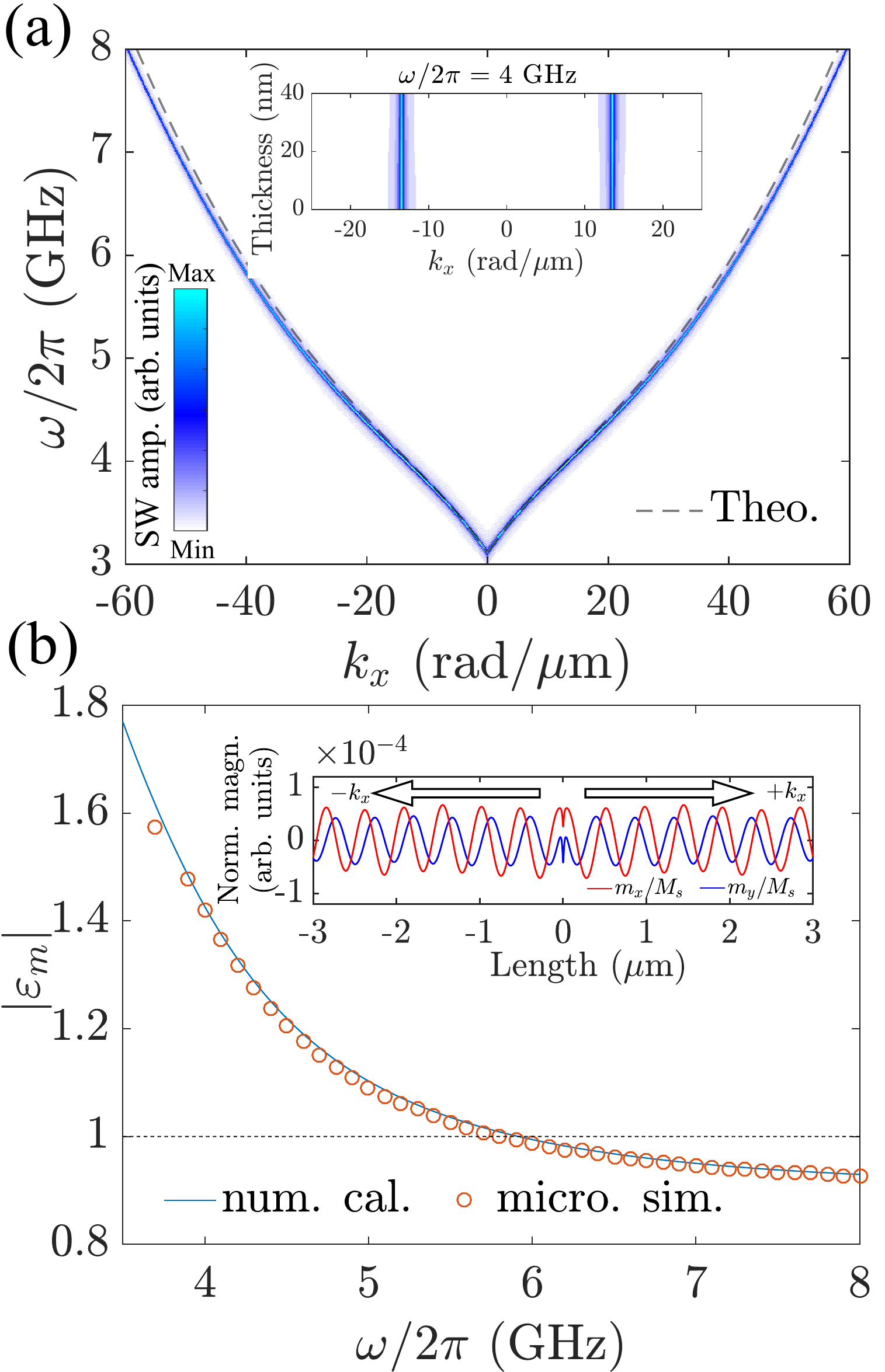}\\
  \caption{(a) SW dispersion relation of the film obtained from micromagnetic simulations. {\color{red}The dashed line represents the theoretical result by Eq. (\ref{DispRela})}. Inset: profiles {\color{red}of SWs with various wavevectors} across the thickness at 4 GHz. (b) Frequency dependence of the dynamic magnetization ellipticity ($|\varepsilon_m|$). The solid curve is from Eq. (\ref{epsilon_m}). Circles are micromagnetic simulations. The dashed line indicates $|\varepsilon_m| = 1$. Inset: Spatial distribution of normalized dynamic magnetization [$m_{x(y)}/M_s$] at 4 GHz at an arbitrary time slot. The blue (red) curves represent the $x$ ($y$) component.} \label{fig1}
\end{figure}

The magnetization dynamics is governed by Landau-Lifshitz-Gilbert (LLG) equation
\begin{equation} \label{LLG}
\frac{\partial {\bf M}}{\partial t} = -\gamma \mu _0{\bf M}\times {\bf H}_{\rm eff} + \frac{\alpha}{M_s} {\bf M} \times \frac{\partial {\bf M}}{\partial t},
\end{equation}
where $\gamma$ is the gyromagnetic ratio, $\mu _0$ is the vacuum permeability, $\alpha\ll1$ is the dimensionless Gilbert damping constant, $M_{s}$ is the saturated magnetization, ${\bf M}={\bf m}+M_s {\bf z}$ is the magnetization with ${\bf m} = m_x {\bf x}+m_y{\bf y}$ the dynamic component, and ${\bf H}_{\rm eff} = {\bf H}_0+{\bf h}_{\rm rf}+{\bf h}_{ex}+{\bf h}_d$ with ${\bf h}_{\rm rf}= h_x  {\bf x} + h_y  {\bf y}$ the microwave field, ${\bf h}_{ex}=(2A_{ex}/\mu _0M_s^2)\nabla ^2{\bf m}$ the exchange field where $A_{ex}$ is the exchange constant, and ${\bf h}_d$ being the dipolar field satisfying the magneto-static Maxwell's equations $\nabla \cdot ({\bf h}_{d}+{\bf m}) = 0$ and $\nabla \times {\bf h}_{d} = 0$. The magnetic parameters of YIG are $M_{s} = 1.48\times 10^5$ A/m, $A_{ex} = 3.1\times 10^{-12}$ J/m, and $\alpha =5\times 10^{-4}$ \cite{Stancil_2009}. The free boundary conditions at the top and bottom surfaces require $\partial m_{x(y)}/\partial y \bigg |_{y=0,-d} =0$ \cite{ZhangPRB2021}. Thus, only the first unpinned mode exists in the low frequency band due to the ultra thin thickness \cite{Mohseni2019,Demokritov2001}, whose profile of dynamic magnetization is uniform across the thickness as shown in the inset of Fig. \ref{fig1}(b). We assume a plane-wave form ${\bf m} = {\bf m}_0 e^{j(\omega t-k_xx)}$ with ${\bf m}_0= m_{x0} {\bf x}+m_{y0} {\bf y}$ and $m_{x(y)} = m_{x0(y0)}e^{j(\omega t-k_xx)}$. Substituting these terms into Eq. (\ref{LLG}) and adopting the linear approximation \cite{Kalinikos1986}, we obtain
\begin{subequations} \label{LLGExpWithRF}
\begin{align}
\label{LLGExpWithRFa}j\omega m_x + (j \alpha \omega + \omega_y) m_y&= \omega_M h_{y}, \\
\label{LLGExpWithRFb}-(j \alpha \omega + \omega_x)m_x + j\omega m_y &= -\omega_M h_{x},
\end{align}
\end{subequations}
where $\omega_{x} = n_x\omega_M+\omega_H+\omega_{ex}$ and $\omega_{y} = n_y\omega_M+\omega_H+\omega_{ex}$, with $\omega_M=\gamma \mu_0M_s$, $\omega_H=\gamma \mu_0 H_0$, and $\omega_{ex}=(2\gamma A/M_s)k_x^2$. The demagnetizing factors $n_x$ and $n_y$ of ${\bf h}_d=-n_xm_x  {\bf x}-n_ym_y  {\bf y}$ are given by (see Appendix \ref{SelfDE} for detailed derivation)
\begin{equation} \label{DemagFac}
n_x=1-n_y=1-\frac{1-e^{-|k_x| d}}{|k_x| d}.
\end{equation}

The non-zero $m_x$ and $m_y$ in Eqs. (\ref{LLGExpWithRF}) requires the determinant
of the coefficient matrix equal to zero, which gives the dispersion relation
\begin{equation} \label{DispRela}
\omega = \sqrt{\omega_{x}\omega_{y}}.
\end{equation}

{\color{red}To verify the theoretical dispersion relation, we perform the simulation using YIG film with 50 $\mu$m long and the excitation with $w = 10$ nm for a broad wave vector range. The excitation is applied using a ``sinc" function ${\bf h}_{\rm rf}(t)=h_0 \sin [\omega _f(t-t_0)]/[\omega _f(t-t_0)]  {\bf x}$ with the cut-off frequency $\omega _f/2\pi=50~\rm {GHz}$, $t_0 = 0.5 ~\rm {ns}$, and $h_0 = 1~\rm {mT}$. The total simulation time is 200 ns, and the results record the dynamic normalized magnetization ($m_y/M_s$) evolution as a function of time and position along $x$ direction. The dispersion relations were obtained through the two-dimensional FFT (2D-FFT) operation on $m_y/M_s$ \cite{Kumar2011}. Figure \ref{fig1}(a) presents a good agreement between the theory and the full micromagnetic simulations. The SW dispersion relation obtained from simulation shows only one band exists in the low frequency range from 3 to 8 GHz, whose profile across the thickness is uniform. Factors $n_x$ and $n_y$ in Eq. (\ref{DemagFac}) are uniquely describing the SW mode with uniform transverse profile [inset of Fig. \ref{fig1}(a)], quite different with those of other SW modes with much higher frequencies \cite{Kalinikos1986}.}
\subsection{ELLIPTICITY}\label{ellipticity}
Solving Eqs. (\ref{LLGExpWithRF}), we obtain
\begin{subequations} \label{LLGExpRF}
\begin{align}
\label{LLGExpRFa}m_x&= \chi_{y}(k_x,\omega) h_{x} +j\kappa(k_x,\omega) h_{y}, \\
\label{LLGExpRFb}m_y&=-j\kappa(k_x,\omega) h_{x} +\chi_{x}(k_x,\omega) h_{y},
\end{align}
\end{subequations}
where
\begin{subequations} \label{Permi}
\begin{align}
\label{Permia}\chi _{x}(k_x,\omega) = -\frac{(\omega_x + j \alpha \omega)\omega_M}{\omega^2-(\omega_x + j \alpha \omega)(\omega_y + j \alpha \omega)}, \\
\label{Permib}\chi _{y}(k_x,\omega) = -\frac{(\omega_y + j \alpha \omega)\omega_M}{\omega^2-(\omega_x + j \alpha \omega)(\omega_y + j \alpha \omega)}, \\
\label{Permic}\kappa(k_x,\omega)  = -\frac{\omega \omega_M}{\omega^2-(\omega_x + j \alpha \omega)(\omega_y + j \alpha \omega)}. &
\end{align}
\end{subequations}
{\color{red}The coefficients $\chi_{x}(k_x,\omega)$, $\chi_{y}(k_x,\omega)$ and $\kappa(k_x,\omega)$ possess the same denominator, whose absolute value takes the minimum when the dispersion relation Eq. (\ref{DispRela}) is satisfied. It means that even though the microwave field comprises multiple wave vector components within $2\pi/w$ \cite{Sushruth2020,Stancil_2009}, only the SWs with $k_x$ and $\omega$ satisfying Eq. (\ref{DispRela}) can be efficiently excited. Substituting Eq. (\ref{DispRela}) into Eqs. (\ref{Permi}) and neglecting the higher-order terms, the magnetic parameters reduce to}
\begin{subequations} \label{LLGExpRF}
\begin{align}
\label{LLGExpRFa}m_x&= \chi_{y} h_{x} +j\kappa h_{y}, \\
\label{LLGExpRFb}m_y&=-j\kappa h_{x} +\chi_{x} h_{y},
\end{align}
\end{subequations}
with
\begin{subequations} \label{PermiApx}
\begin{align}
\label{PermiApxa}\chi _{x}& = -\frac{j\omega_x \omega_M}{\alpha (\omega_x + \omega_y )\sqrt {\omega_x  \omega_y}}, \\
\label{PermiApxb}\chi _{y}& = -\frac{j\omega_y \omega_M}{\alpha (\omega_x + \omega_y )\sqrt {\omega_x  \omega_y}}, \\
\label{PermiApxc}\kappa & = -\frac{j\omega_M}{\alpha (\omega_x + \omega_y )}.
\end{align}
\end{subequations}
We obtain the ratio between the $x$ and $y$ components of the dynamic magnetization as
\begin{equation} \label{epsilon_m}
\varepsilon_m=\frac{m_x}{m_y}=j\sqrt{\frac{\omega_{y}}{\omega_{x}}}
=j\sqrt{\frac{n_y\omega_M+\omega_H+\omega_{ex}}{n_x\omega_M+\omega_H+\omega_{ex}}}.
\end{equation}

Equation (\ref{epsilon_m}) delivers following features of spin precessions: (i) the imaginary unit $j$ in $\varepsilon_m$ implies spin precessions in ferromagnetic films are always right-hand polarized, as shown in the inset of Fig. \ref{fig1}(b) where $m_y$ drops behind $m_x$ for $1/4$ wavelength regardless of their propagating directions. (ii) The $\varepsilon_m$ is irrelevant to the amplitudes or phases of $h_x$ and $h_y$. In the exchange limit of $k_x\rightarrow\infty$, $\varepsilon_m\rightarrow j$ indicates the SWs are perfectly right-circularly polarized \cite{YuSpringer2021}. Meanwhile, in the dipolar-exchange region where the dipolar effect is comparable to exchange interaction, $\varepsilon_m$ varies with factors $n_x$ and $n_y$, which rely on $(\omega, k_x)$, as shown in Fig. \ref{fig1}(b).
\begin{figure}[htbp!]
  \centering
  \includegraphics[width=0.48\textwidth]{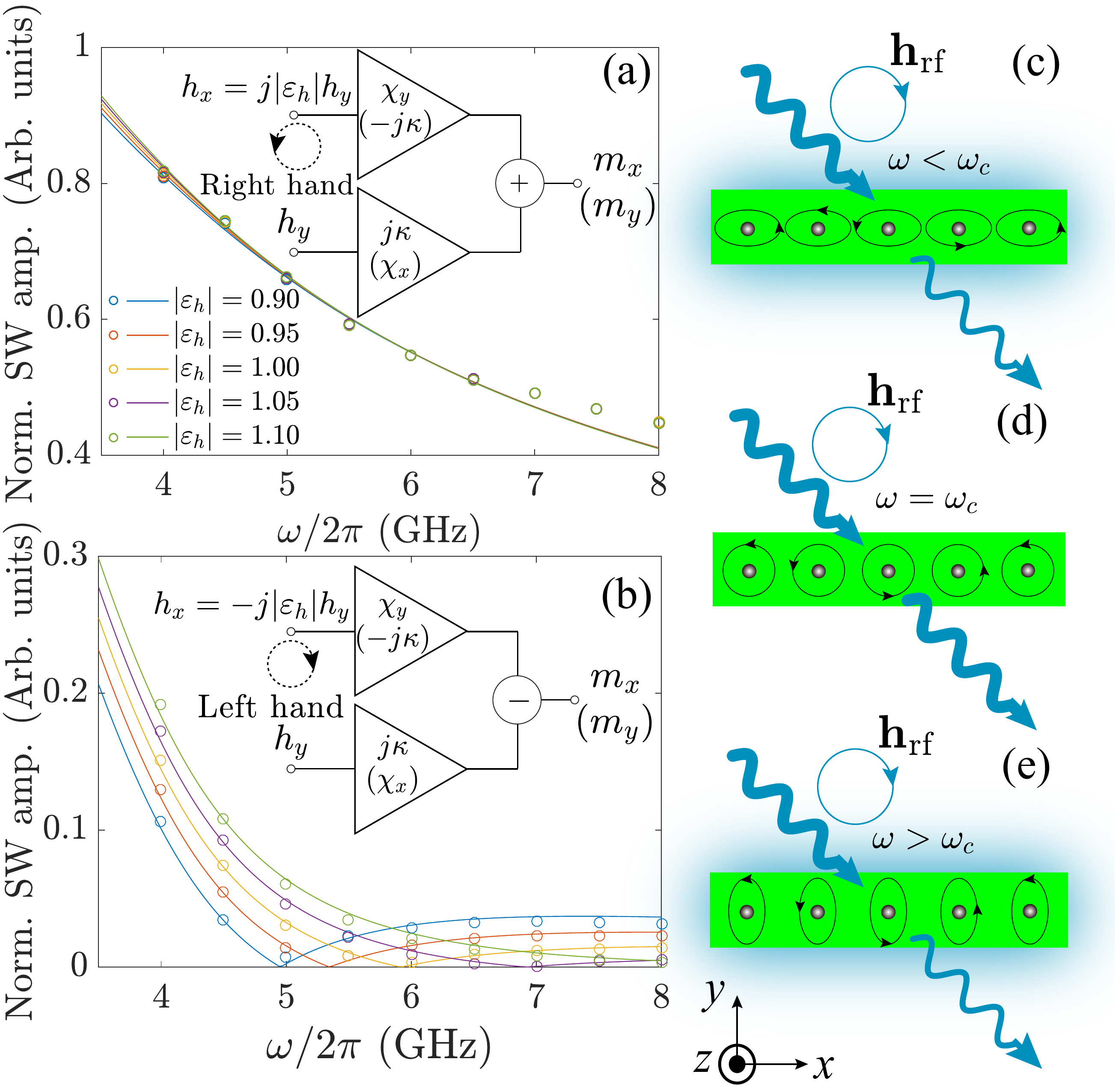}\\
  \caption{ SW amplitudes ($|{\bf m}|$) normalized by the maximal value excited by (a) the right- and (b) left-hand chiral microwave fields with the same power density but different ellipticities ranging from 0.9 to 1.1. Solid curves are calculated based on Eqs. (\ref{LLGExpRF}). Insets in (a) and (b) depict the schematics of constructive and destructive superposition of the contributions of $h_x$ and $h_y$, in analog to the common and different signals in differential amplifiers, respectively. Symbols are from micromagnetic simulations. Illustrations of the left-hand chiral photon-magnon conversion (c) below, (d) at and (e) above $\omega_c$ {\color{red}, respectively}. The blue wavy arrays and circles {\color{red}represent} the {\color{red}microwave} fields with thickness indicating the intensity, {\color{red} where the blue arrowed circles represent the chirality. The blue glowing backgrounds indicate the converted microwave energy}.  The dots and circles represent the spins and their precession cones, respectively.} \label{fig2}
\end{figure}

\subsection{INTENSITY SPECTRA}\label{ChiExci}
Below, we investigate SW amplitudes $|{\bf m}| =\sqrt{m_x^2+m_y^2}$ dependence on the microwave field chiralities. Specifically, we inspect the typical cases that  $\varepsilon_h = h_x/h_y$ is purely imaginary where {\color{red}$h_x = j|\varepsilon_h|h_y$} and {\color{red}$h_x = -j|\varepsilon_h|h_y$} represent respectively the right- and left-handed polarization with $\left |\varepsilon_h \right |$ being their ellipticity. In micromagnetic simulations, the excitation is applied using the function ${\bf h}_{\rm rf}(t) = h_{x0}\sin (\omega t) {\bf x} + h_{y0} \sin (\omega t\pm \pi/2) {\bf y}$, with ``+" (``$-$") for the left (right) hand polarization. We fix  $h_0 = \sqrt{h_{x0}^2+h_{y0}^2}=0.1~{\rm mT}$ to ensure the same RF power density with different ellipticities. The results record the dynamic normalized magnetization ($m_x/M_s$ and $m_y/M_s$) evolution as a function of time and space. The amplitude spectra of the SWs excited by the right- and left-hand polarized microwaves with $|\varepsilon_h|$ ranging from 0.9 to 1.1 are plotted in Figs. \ref{fig2}(a) and  \ref{fig2}(b), respectively.

The chiral excitation of SWs possesses the following features. Firstly, the complex parameters $\chi_{x}$, $\chi_{y}$ and $\kappa$ expressed by Eqs. (\ref{PermiApx}) take the same phase factor. Therefore, Eqs.  (\ref{LLGExpRF}) manifest that the contribution of $h_y$ to ${\bf m}$ is delayed by the phase of $\pi/2$ compared to that of $h_x$. Consequently, they are superposed destructively (constructively) in the case of left- (right-) hand excitation. Simulation results confirm this point that the left-hand excited SW intensities are weaker than its right-hand counterpart. We thus introduce an analog to the differential amplifier in electronic systems, in which the dual inputs are separately amplified, subtracted (added) and output as different (common) mode signals \cite{Yawale2022}, as illustrated by the insets in Figs. \ref{fig2}(a) and \ref{fig2}(b). The dual inputs, amplifying factors and outputs are in comparison with ${\bf h}_{\rm rf}$, the complex parameters ($\chi_{x}$, $\chi_{y}$ or $\kappa$) and ${\bf m}$, respectively. And $\varepsilon_h$ reflects the ratio between the dual inputs. {\color{red} Then we obtain
\begin{equation} \label{DiffAmpMod}
\left[
 \begin{aligned}
 m_{x} \\
 m_{y}
 \end{aligned}
\right]
=
\left[
 \begin{aligned}
jh_y(\kappa\pm|\varepsilon_h|\chi_y)\\
h_y(\chi_x\pm|\varepsilon_h|\kappa)
 \end{aligned}
\right],
\end{equation}
with ``+'' and ``$-$'' for the results of left- and right-handed excitations respectively, which certify the validation of the differential amplifier model.
}

Secondly, using the differential amplifier model, it can be explained that the left-hand excited SW spectra are much more sensitive to the variation of $\varepsilon_h$ than the right-hand excited ones since the differential (common) mode signal is sensitive (irresponsive) to the tiny variation ($\varepsilon _h$) of the dual inputs ($h_x$ and $h_y$). It is observed that the curves describing different $\varepsilon_h$ in Fig. \ref{fig2}(a) are almost merged, while those in Fig. \ref{fig2}(b) are well separated. {\color{red}Moreover, the intensity of left-hand excited SWs is much weaker than that of their right-hand counterparts. Especially the former drops to almost one tenth of the latter at high frequencies.}

Lastly, even though the two pairs of amplifying factors [($\chi_{y}$, $\kappa$) and ($\kappa$, $\chi_{x}$)] for the outputs $m_x$ and $m_y$ are different, their ratios are both $1/|\varepsilon_m|$. Mathematically, we can substitute Eqs. (\ref{PermiApx}) and Eq. (\ref{epsilon_m}) into Eqs. (\ref{LLGExpRF}), and obtain {\color{red} a more generalized expression for $\bf m$}
\begin{equation} \label{Chiral_Exci}
\left[
 \begin{aligned}
 m_{x} \\
 m_{y}
 \end{aligned}
\right]
=-\frac{\omega_m h_y\left(\varepsilon_h \varepsilon_m -1 \right)}{\alpha\left(\omega_x+\omega_y\right)}
\left[
 \begin{aligned}
 &1 \\
 1&/\varepsilon_m
 \end{aligned}
\right].
\end{equation}
It suggests a compensation frequency ($\omega_c$) when $\varepsilon_h\varepsilon_m=1$. Equation (\ref{epsilon_m}) indicates $\Im\left(\varepsilon_m\right)>0$, therefore only the the left-hand microwaves with $\Im\left(\varepsilon_h\right)<0$ support $\omega_c$, at (below and above) which the microwaves with any intensities are unable (able) to excite any SWs, as illustrated by Figs. \ref{fig2}(c), \ref{fig2}(d) and \ref{fig2}(e). The equality of the ratios is also the prerequisite for treating SWs as scalar variables in previous researches \cite{VasilievJAP2007,KostylevPRB2007,DemokritovPRL2004}. This finding broadens the selection rule for photon-magnon conversion, which is instructive to the chiral magneto-optic and -acoustic effects \cite{ZhangPRAppl2020,XuSciAdv2020}. However, $\omega_c$ cannot exist for arbitrary $\varepsilon_h$ because $|\varepsilon_m|$ given by Eq. (\ref{epsilon_m}) can only take values from 0.91 to 2.14 for the present model parameters. Consequently, $\omega_c$ can only emerge with $\varepsilon_h$ in the range from 0.47 to 1.09. Analytical and numerical results indeed confirm this point that the curve for $|\varepsilon_h| = 1.1$ (the green one) in Fig. \ref{fig2}(b) cannot intersect with $x$ axis.

\section{STRATEGY FOR NONRECIPROCAL SPIN WAVES}\label{Nonreciprocal}

\begin{figure}[htbp!]
  \centering
  \includegraphics[width=0.40\textwidth]{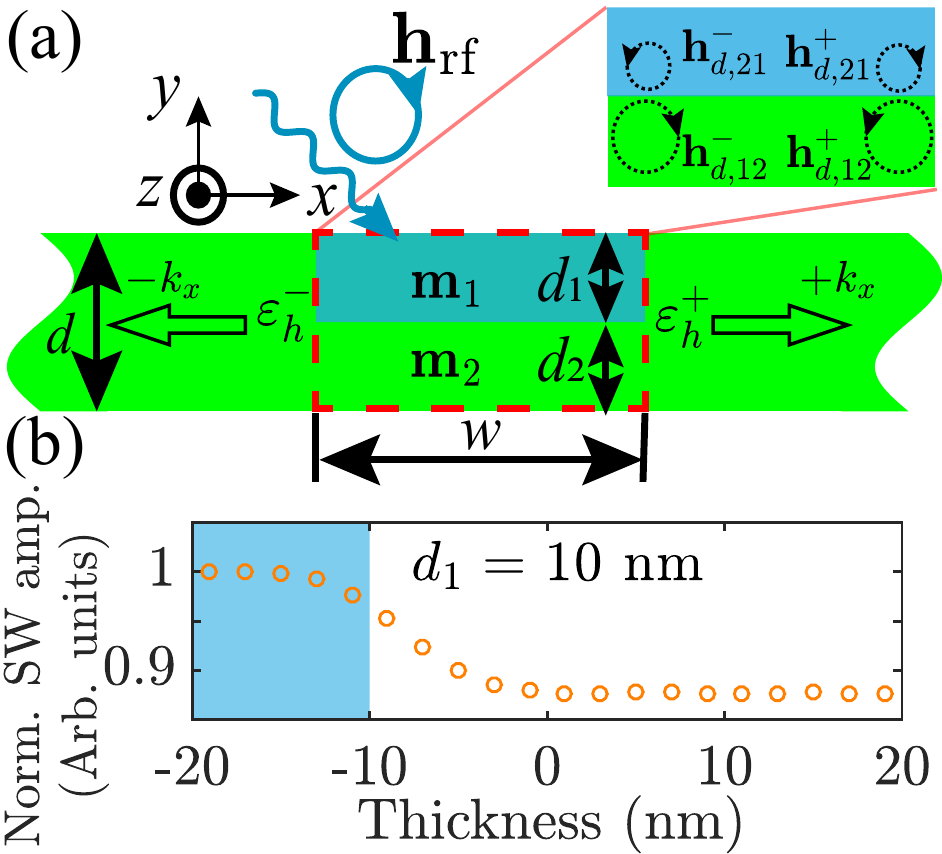}\\
  \caption{(a) The schematic for the non-reciprocal SW excitation using left-hand chiral microwave field applied on the top part of the film in the patched blue area. Inset shows the precession cones of the mutual dipolar fields induced by the forward and backward propagating SWs in each layer, with the amplitude indicated by the radius. (b) Simulated SW amplitudes in red box for $d_1 = 10$ nm at 4.6 GHz.} \label{fig3}
\end{figure}

The above discussions indicate that using the left-hand excitation is essential to generate the non-reciprocal SWs {\color{red}due to its high sensitivity of the spectra to $\varepsilon_h$. We only need to slightly alter the ellipticity ($\varepsilon_h^+$ and $\varepsilon_h^-$) of microwave fields for exciting the forward and backward propagating SWs with quite different spectra (superscripts ``$+$'' and ``$-$'' are used to label the forward and backward parameters, respectively, same hereinafter). In comparison, the right-hand exciting cases require the $\varepsilon_h^+$ and $\varepsilon_h^-$ to be dramatically varied for substantially different spectra. Hence, the left-handed excitation brings remarkable convenience for designing the method on nonreciprocity.} One critical technique is to differentiate $\varepsilon_h^+$ and $\varepsilon_h^-$. In the multi-layer structure, the dynamic mutual dipolar effect between layers has been demonstrated to be directionally dependent \cite{Henry2016}. So, one natural issue arises if we can introduce the mutual dipolar field combined with ${\bf h}_{\rm rf}$ to differentiate $\varepsilon_h^+$ and $\varepsilon_h^-$. Here, we investigate the method by applying microwave field unevenly across the film thickness.  This idea is in contrast to preceding works, where {\color{red}additional micro-magnets outside YIG films are indispensable to serve as the spin wave source } and the effective exciting polarizations are simply circular with directionally opposite chirality, resulting in the nonreciprocity \cite{WangNR2020,ChenPRB2019,YuPRB2019,YuPRL2019,YuSpringer2021}. For simplification, ${\bf h}_{\rm rf}$ is uniformly applied only on the top part of the film with thickness $d_1$ {\color{red}and width $w$}, as shown in Fig. \ref{fig3}(a). The SW characteristics dependence on the excitation is investigated by varying $d_1$. {\color{red}The spin wave information in the exciting area along thickness is extracted by calculating $|\bf {m}|$ at every mesh grid, averaged in the exciting area with width $w$ and normalized by the maximal value.} We estimate the different dynamic magnetizations (${\bf m}_1= m_{x,1}  {\bf x}+m_{y,1}  {\bf y}$ and ${\bf m}_2= m_{x,2}  {\bf x}+m_{y,2}  {\bf y}$) along $d_1$ and $d_2=d-d_1$ that introduces the mutual demagnetizing field, whose simulated SW amplitudes are shown in Fig. \ref{fig3}(b). Even though the minor inhomogeneity appears at the interface, they are approximated to be transversely uniform in the following analysis. The part in dashed red box taking bilayer structure is regarded as the SW source. In this case, the dipolar fields are composed of two components: the self demagnetizing field ${\bf h}_{d,p}=-n_{x,p} m_{x,p}  {\bf x}-n_{y,p} m_{y,p}  {\bf y}$ where $n_{x(y),p}$ is given by Eq. (\ref{DemagFac}) with $n_{x(y)} \rightarrow n_{x(y),p}$ and $d\rightarrow d_{p}$, and the mutual demagnetizing field ${\bf h}_{d,pq}=h_{d,x,pq}  {\bf x}+h_{d,y,pq}  {\bf y}$ [($p,q$) = (1,2) or (2,1)]. Here $h_{d,x(y),pq}$ satisfy the following identity (see Appendix \ref{MutDE} for detailed derivation)
\begin{equation} \label{DipFldMutil}
\left[
 \begin{aligned}
 h_{d,x,pq} \\
 h_{d,y,pq}
 \end{aligned}
\right]=-n_{pq}
\left[
\begin{matrix}
     1 & j\sign{k_{x}}(q-p) \\
     j\sign{k_{x}}(q-p) & -1
\end{matrix}
\right]
\left[
 \begin{aligned}
 m_{x,p} \\
 m_{y,p}
 \end{aligned}
\right],
\end{equation}
with
\begin{equation} \label{GenDemFac}
n_{pq}=
\frac{\left(1-e^{-|k_{x}|d_p}\right)\left(1-e^{-|k_{x}|d_q}\right)}{2|k_{x}|d_q}.
\end{equation}
The ${\bf h}_{d,pq}$ (${\bf h}_{d,p}$) is directionally dependent (independent) according to Eqs. (\ref{DipFldMutil}) [Eq. (\ref{DemagFac})]. Hence, ${\bf h}_{d,pq}$ rather than ${\bf h}_{d,p}$ contributes to the non-reciprocity. In addition, ${\bf h}_{d,21}^{+}$ (${\bf h}_{d,21}^{-}$) and ${\bf h}_{d,12}^{+}$ (${\bf h}_{d,12}^{-}$) are contrarily circular polarized with different intensities, as sketched in the inset of Fig. \ref{fig3}(a) [as proved by Eqs. (\ref{DeMut}) in Appendix \ref{MutDE}]. The net effective mutual field ${\bf h}_{d,{\rm mut}}$ on the entire film is therefore given by
\begin{widetext}
\begin{equation} \label{NetDeMut}
\begin{aligned}
{\bf h}_{d,{\rm mut}}=\frac{{\bf h}_{d,12}d_2+{\bf h}_{d,21}d_1}{d}
=-n_{\rm mut}\left\{
 \begin{aligned}
 \left[(m_{x,1}+m_{x,2})+j\sign {k_x}(m_{y,1}-m_{y,2})\right]   {\bf x}
 +\left[j\sign {k_x}(m_{x,1}-m_{x,2})-(m_{y,1}+m_{y,2})\right]  {\bf y}
 \end{aligned}
\right\}.
\end{aligned}
\end{equation}
\end{widetext}
with
\begin{equation} \label{NetDeMutFac}
\begin{aligned}
n_{\rm mut} = \frac{\left(1-e^{-|k_{x}|d_1}\right)\left(1-e^{-|k_{x}|d_2}\right)}{2|k_x|d}.
\end{aligned}
\end{equation}

\begin{figure*}[htbp!]
  \centering
  \includegraphics[width=0.96\textwidth]{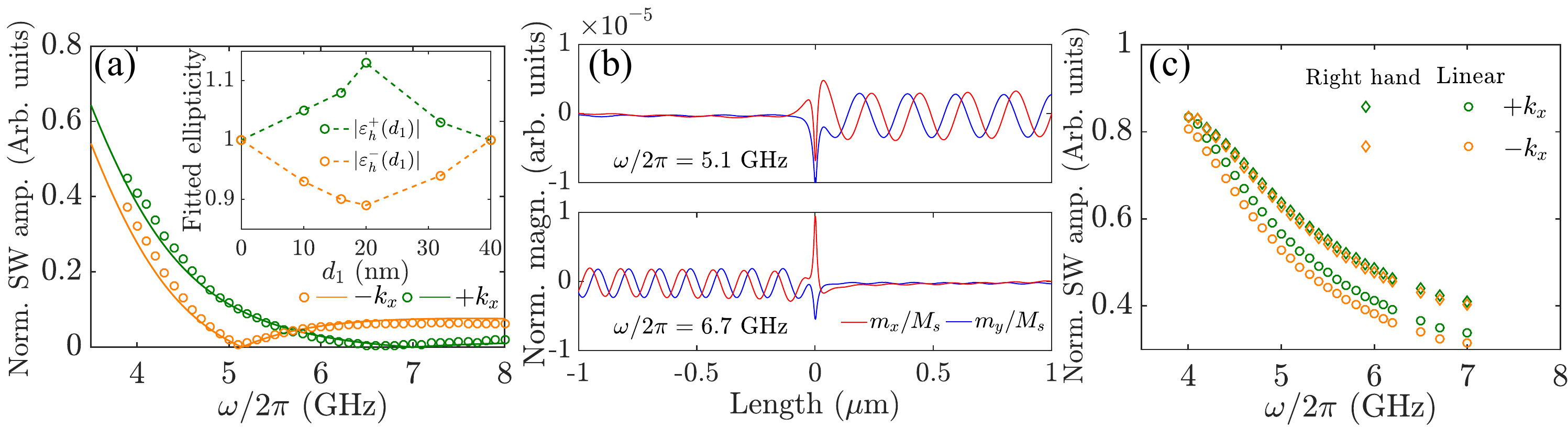}\\
  \caption{(a) Spectra of the forward (green) and backward (orange) SWs amplitudes with excitation depths $d_1=20~{\rm nm}$. The intensities are normalized with the maximal value. Symbols are numerical simulations and the curves are fitting results. The inset of (a) shows the fitted $\varepsilon_h^{+}$ and $\varepsilon_h^{-}$ dependence on $d_1$. (b) Simulated $m_{x}/M_s$ and $m_{y}/M_s$ distribution at two compensation frequencies, $\omega/2\pi=5.1~{\rm and ~6.7~ GHz}$, respectively. (c)  Simulated forward (green symbols) and backward (orange symbols) SW spectra under the right-hand (diamonds) and linear (circles) excitations unevenly applied across the film thickness ($d_1 = 10$ nm).} \label{fig4}
\end{figure*}

Following conclusions can thus be drawn. Firstly, the non-reciprocity disappear if $d_1 = 0$ or $d_2 = 0$, which causes $n_{\rm mut}=0$ and ${\bf h}_{d,{\rm mut}}=0$. It was confirmed that the SWs propagating along opposite directions share the same amplitude with the uniform excitation across the thickness, as shown in Fig. \ref{fig1}(b).
Secondly, since ${\bf h}_{d,{\rm mut}}$ is determined by ${\bf m}_{1}$ and ${\bf m}_{2}$, its role is to tune the two gains in the differential amplifier [insets of Figs. \ref{fig2}(a) and \ref{fig2}(b)], equivalent to varying $\varepsilon_h$ of the input microwave ${\bf h}_{\rm rf}$. As the variation of $\varepsilon_h$ is directional with ${\bf h}_{d,{\rm mut}}$, the intensity spectra are well separated for the forward and backward SWs, as plotted in Fig. \ref{fig4} (a). Even though ${\bf h}_{d,{\rm mut}}$ is frequency dependent, simulation results can still be well fitted using Eqs. (\ref{LLGExpRF}) and $h_x = \varepsilon_h^{+(-)}(d_1) h_y$ with $\omega_c$ satisfying $\varepsilon_h^{+(-)}(d_1)\varepsilon_m = 1$, where $\varepsilon_h^{+(-)}(d_1)$ is the effective ellipticity to be determined. The fitted $|\varepsilon_h^{+(-)}(d_1)|$ is plotted in the inset of Fig. \ref{fig4}(a) {\color{red}, where the goodness of all fittings is greater than $91\%$}. Representatively, we obtain $|\varepsilon_h^{+}(10~{\rm nm})|=1.05$ and $|\varepsilon_h^{-}(10~{\rm nm})|=0.93$, corresponding to $\omega_c/2\pi = 5.1$ and 6.7 GHz, with the dynamic magnetization presented in upper and lower panels of Fig. \ref{fig4}(b), respectively. {\color{red}Sacrificing the efficiency of excitations with the amplitude one order lower than that in the inset of Fig. \ref{fig1}(b), we can obtain theoretically switchable non-reciprocities and one hundred percentage (perfect) unidirectionality. This is advantageous over many other strategies \cite{XuSciAdv2020}.}
Thirdly, the difference between $\varepsilon_h^{+}$ and $\varepsilon_h^{-}$ and the separation of the forward and backward SW intensity spectra approaches the maximum at $d_1=d_2=d/2$, meeting the maximal value condition of $n_{\text{mut}}$ in Eq. (\ref{NetDeMutFac}). {\color{red} Notwithstanding, the value of $|\varepsilon_h^{+}(d_1 = 20 {\rm nm})|=1.13$ exceeds the range from 0.47 to 1.09. It} indicates that {\color{red} no $\omega_c$ would present in the forward SW spectra as discussed in Sec. \ref{ChiExci} . Consequently,} the perfect backward SW propagation without any forward SW cannot be achieved.

Lastly, for completeness, we perform the simulations applying right-hand and linear microwave on the top part of the films with $d_1 = 10$ nm. The forward and backward SW spectra are presented in Fig. \ref{fig4}(c). {\color{red}Following features are observed. (i) Both spectra} are not well separated, indicating {\color{red}that $\bf h_{\rm {rf}}$ can induce non-reciprocity, but the effect is not significant. It can be understood in this configuration since ferromagnetic films are much thinner than the spin wavelengths \cite{YuPRB2019,YuPRL2019}. (ii)} The forward SWs are always stronger than the backward ones in the whole frequency band, implying that DE mechanism induced nonreciprocity cannot be switched by tuning frequencies since it is merely dependent on the surface normal and static magnetization directions \cite{Kwon2016,An2013}. {\color{red}In conclusion, the switchable and perfect non-reciprocities do not appear in the spectra of right-hand and linear excitation, reconfirming that they are the unique features of the left-hand excited SWs.}

\begin{figure}[htbp!]
  \centering
  \includegraphics[width=0.4\textwidth]{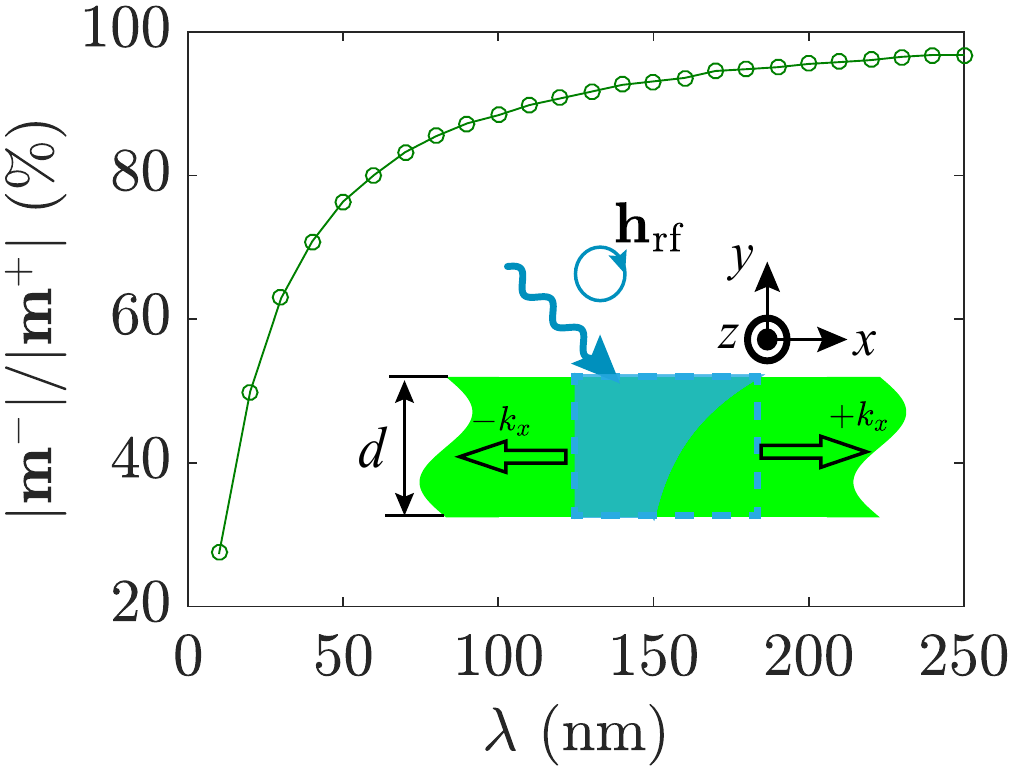}\\
  \caption{Simulated $|\bf{m}^-|/|\bf{m}^+|$ dependence on the decay length of the exponential profiled left-handed excitations at 4.6 GHz. The line connecting the symbols guide the trend. The inset schematically shows the simulated structure with $d = 40$ nm. The microwave field is applied on dashed box area with exponential intensity profile indicated by the patched blue area. } \label{fig5}
\end{figure}

Finally, we note that the key for exciting non-reciprocal SWs is the introduction of non-zero ${\bf h}_{d,\rm mut}$, induced by the asymmetrically distributed ${\bf m}$ across thickness, which can be simply excited by uneven profiled ${\bf h}_{\rm rf}$. Such fields can be generated by the resonant spin nano-oscillators with various structures, like nano-disks \cite{Demidov2012} or  nano-wires \cite{Safranski2017}{\color{red}, which are less sharp than that proposed in the above analysis}. {\color{red} Even so, ${\bf h}_{\rm rf} $ with more gradual uneven profiles can also excite non-reciprocal SWs. To verify this point, we perform simulations using left-handed ${\bf h}_{\rm rf}$ with exponential profiles and $|\varepsilon_h| = 1$, as shown in the inset of Fig. \ref{fig5}. The intensity dependence on the thickness is described by $h_0(y) = h_0(\lambda)e^{-y/\lambda}$, where $h_0(\lambda)$ is determined by $\int_{-d}^{0} h_0(y)dy = h_0d ~ (h_0 = 0.1 \rm{mT})$ to ensure the same power intensity. The ratio $|\bf{m}^-|/|\bf{m}^+|$ depending on the decay length $\lambda$ is plotted in Fig. \ref{fig5}. When $\lambda$ is shorter than $d$, the ratio is observed to rise dramatically with the increase of $\lambda$ due to the rapid decrease of the uneven degree of the exciting field. It generally converges to $100 \%$ with $\lambda\rightarrow\infty$, corresponding to the case without the non-reciprocity.} Furthermore, the switched non-reciprocity at two compensation frequencies contributes additional methodology for magnonic frequency division multiplexing, broadening the strategy for designing magnonic circuits \cite{ZhangAPL2019}.

\section{DISCUSSION AND CONCLUSION}\label{Conclusion}

In summary, we investigated the propagating dipolar-exchange SWs excited by chiral microwaves in ferromagnetic thin films. We showed that the left-hand microwave can excite non-reciprocal SWs in the condition of ellipticity mismatch. When the left-hand microwave is unevenly applied across the film thickness, we observed a SW chirality switching by tuning the microwave frequency. Our findings shine a new light on the photon-magnon conversion and pave the way toward engineering the nano-scaled chiral microwave field for the realization of the diode-like functionalities in magnonics.

\section{ACKNOWLEDGEMENTS}
\begin{acknowledgments}\label{Acknowledgments}
We thank Y. Henry for helpful discussions. This work was supported by the National Natural Science Foundation of China (NSFC) (Grants No. 12074057, No. 11604041, and No. 11704060). Z.Z. acknowledges the financial support of the China Postdoctoral Science Foundation under Grant No. 2020M673180. Z.W. was supported by the China Postdoctoral Science Foundation under Grant No. 2019M653063 and the NSFC (Grant No. 12204089). Z.-X.L. acknowledges financial support from the China Postdoctoral Science Foundation (Grant No. 2019M663461) and the NSFC (Grant No. 11904048).
\end{acknowledgments}

\begin{appendix}

\section*{APPENDIX}
We investigate the dipolar effects induced by the SWs propagating in the ultra thin magnetic film. The dipolar field in the whole space is calculated. The self and mutual demagnetizing factors are figured out in Sec. \ref{SelfDE} and Sec. \ref{MutDE}, respectively.
We considered a magnetic film extended infinitely along $x$ and $z$ directions, located from $y = -d$ to 0 and labelled as $L_i$. The SWs takes the form ${\bf m}_i = {\bf m}_{0,i} e^{j(\omega t-k_xx)}=m_{x,i} {\bf x}+m_{y,i} {\bf y}$ with ${\bf m}_{0,i}= m_{x0,i} {\bf x}+m_{y0,i} {\bf y}$. The dynamic magnetization ${\bf m}_i$ and the correspondingly induced dipolar field ${\bf h}_{d,i}$ satisfy the magnetostatic equations
\begin{subequations} \label{MaxEqs}
\begin{align}
\label{MaxEqsa}\nabla \cdot ({\bf h}_{d,i}+{\bf m}_i) = 0, \\
\label{MaxEqsb}\nabla \times {\bf h}_{d,i} = 0. &
\end{align}
\end{subequations}
Introducing the scale potential $\psi_{m,i}$, we have
\begin{equation} \label{Magn_Pot}
{\bf h}_{d,i}=-\nabla \psi_{m,i}.
\end{equation}
Then Eq. (\ref{MaxEqsa}) becomes Poisson equation
\begin{equation} \label{PoiEq}
\nabla^2 \psi_{m,i} = -\rho_{i},
\end{equation}
where $\rho_{i}$ is the effective magnetic-charge density, given as
\begin{equation} \label{MagChg}
\rho_{i} = -\nabla \cdot {\bf m}_i.
\end{equation}
One crucial step is to find the solution of $\psi_{m,i}$ in Eq. (\ref{PoiEq}). We note that there are two contributions to $\psi_{m,i}$ in magnetic materials: effective \emph{volume} magnetic-charge density $\rho_{m,i}$ and effective \emph{surface} magnetic-charge density $\sigma_{m,i}$ \cite{Jackson1962}.

\begin{figure}
  \centering
  \includegraphics[width=0.48\textwidth]{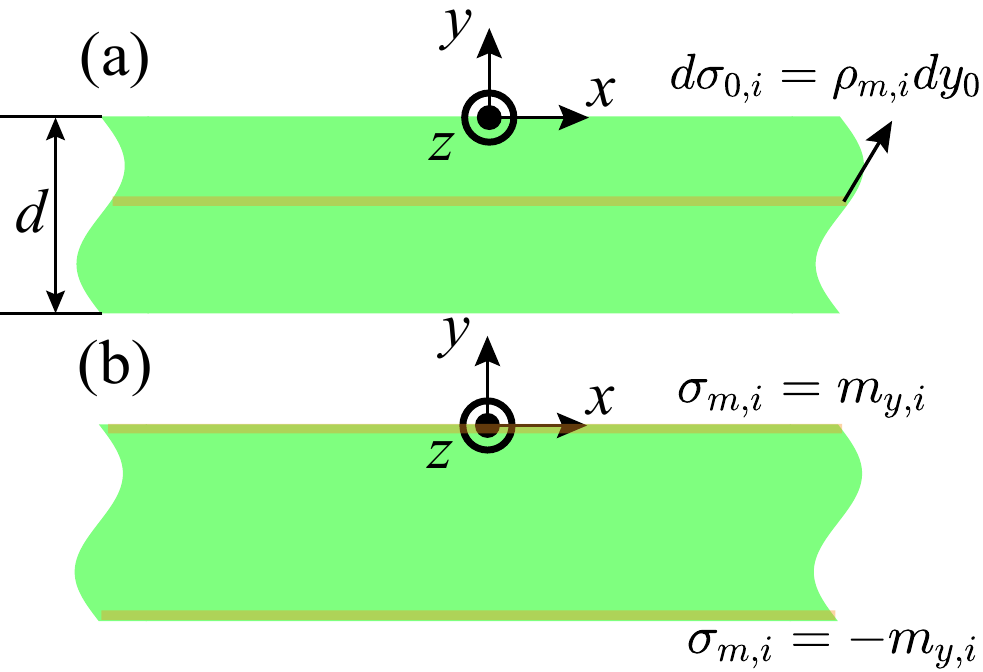}\\
  \caption{Schematics of the effective (a) \emph{volume} and (b) \emph{surface} magnetic-charges. The patched yellow parts represent the differential elements} \label{SMfig1}
\end{figure}

Firstly, we calculate the contribution of $\rho_{m,i}$. Inside the film, $\rho_{m,i}=-\nabla \cdot {\bf m}_i=jk_xm_{x,i}$ is induced by the $x$ component of $m_i$ \cite{Henry2016}. To begin with, we consider a tiny sheet of film located at position $y = y_0$ with thickness $dy_0$, whose surface magnetic charge density is $\sigma_{0,i} = \rho_{m, i}dy_0$ [see Fig. \ref{SMfig1}(a)]. The magnetostatic potential $\psi_{m, i}(\sigma_{0,i}, y_0, {\bf r}, t)$ induced by $\sigma_{0,i}$ is periodic (evanescent) along $x$ ($y$) direction, while its maximum locates at $y = y_0$ and satisfy Laplace equation $\nabla^2\psi_{m, i}(\sigma_{0,i}, y_0, {\bf r}, t) = 0$ \cite{Bailleul2011}. Then the solution can be expressed as
\begin{equation} \label{PotEle}
\psi_{m, i}(\sigma_{0,i}, y_0, {\bf r}, t)
 = \psi_{m0,i}(\sigma_{0,i})e^{-|k_x(y-y_0)|}e^{j(\omega t-k_xx)}.
\end{equation}
The next important step is to find out the value of $\psi_{m0,i}$. Note that the boundary condition (continuity of $B_y$) of the tiny sheet is given as
\begin{equation} \label{BdC}
h_{y,i}(\sigma_{0,i},y_0^+,{\bf r},t)-h_{y,i}(\sigma_{0,i},y_0^-,{\bf r},t) = \sigma_{0,i}.
\end{equation}
Using Eq. (\ref{Magn_Pot}), we have
\begin{equation} \label{DipField}
\begin{aligned}
h_{y,i}(\sigma_{0,i},y_0,{\bf r},t)
 & = -\frac{\partial}{\partial y} \psi_{m,i}(\sigma_{0,i},y_0,{\bf r},t)\\
 & = \left \{
 \begin{aligned}
 |k_x|\psi_{m0,i}e^{|k_x|(y-y_0)}e^{j(\omega t-k_xx)},&y\geq y_0\\
 |k_x|\psi_{m0,i}e^{-|k_x|(y-y_0)}e^{j(\omega t-k_xx)},&y<y_0\\
 \end{aligned}
 \right .
\end{aligned}
\end{equation}
Therefore, we have
\begin{equation} \label{SolPot}
2|k_x|\psi_{m0,i}(\sigma_{0,i})e^{j(\omega t-k_xx)}=\sigma_{0,i}.
\end{equation}
The magneto-static potential induced by the sheet at $y = y_0$ can be expressed as
\begin{equation} \label{SolvedPot}
\psi_{m,i}(\sigma_{0,i},y_0,{\bf r},t)
=\frac{j\sign {k_x}m_{x0,i}e^{j(\omega t-k_xx)}}{2}e^{-|k_x(y-y_0)|}dy_0.
\end{equation}
Correspondingly, the dipolar magnetic field ${\bf h}_{d,i}(\sigma_{0,i},y_0,{\bf r},t)$ derived from $\psi_{m,i}(\sigma_{0,i},y_0,{\bf r},t)$ is given as
\begin{equation} \label{SolvedDipFld}
\begin{aligned}
& {\bf h}_{d,i}(\sigma_{0,i},y_0,{\bf r},t)\\
= &\frac{j\sign {k_x}m_{x,i}}{2}e^{-|k_x(y-y_0)|}
\left [-\sign{k_x} {\bf x}+j\sign{y-y_0} {\bf y}\right ]dy_0.
\end{aligned}
\end{equation}
The dipolar field ${\bf h}_{d,i}(\rho_{m,i},{\bf r},t)$ induced by $\rho_{m,i}$ at any position ${\bf r}$ is given
\begin{widetext}
\begin{equation} \label{AnyVolFld}
\begin{aligned}
{\bf h}_{d,i}(\rho_{m,i},{\bf r},t)& =\frac{1}{2}\int_{-d}^{0}j\sign {k_x}m_{x,i}e^{-|k_x(y-y_0)|}\left [-\sign{k_x}  {\bf x}+j\sign{y-y_0} {\bf y}\right ]dy_0\\
 &= \left \{
\begin{aligned}
& \frac{m_{x,i}}{2}e^{-|k_x|y}\left (1-e^{-|k_x|d}\right )\left [- {\bf x}+j\sign {k_x} {\bf y}\right ],&y\geq0\\
& -\frac{m_{x,i}}{2}\left [2-e^{-|k_x|(y+d)}-e^{|k_x|y}\right ] {\bf x}+\frac{jm_{x,i}}{2}\sign {k_x}\left [e^{|k_x|y}-e^{-|k_x|(y+d)}\right] {\bf y},&-d\leq y<0\\
&  \frac{m_{x,i}}{2}e^{|k_x|y}\left (e^{|k_x|d}-1\right )\left [- {\bf x}-j\sign {k_x} {\bf y}\right ],&y<-d\\
\end{aligned}
\right.
\end{aligned}
\end{equation}
\end{widetext}
Next, we calculate the contribution of $\sigma_{m,i} = {\bf m}_{i}\cdot {\bf  {n}}$, located only at the position $y = 0$ and $y = -d$ with ${\bf  {n}}$ the unit vector normal to the surface. They are equal to $m_{y, i} = m_{y0, i} e^{j(\omega t-k_xx)}$ and $-m_{y, i}$, where the minus sign comes from the opposite directions of the top and bottom surfaces.Following the steps from Eqs. (\ref{PotEle}) to (\ref{SolvedPot}), we obtain the magneto-static potential induced by $\sigma_{m,i}$
\begin{subequations} \label{SurfPot}
\begin{align}
\label{SurfPota}\psi_{m,i}(\sigma_{m,i},0,{\bf r},t) &= \frac{m_{y0,i}}{2|k_x|}e^{j(\omega t-k_xx)}e^{-|k_xy|}, \\
\label{SurfPotb}\psi_{m,i}(\sigma_{m,i},-d,{\bf r},t) &= -\frac{m_{y0,i}}{2|k_x|}e^{j(\omega t-k_xx)}e^{-|k_x(y+d)|}.
\end{align}
\end{subequations}
The dipolar field ${\bf h}_{d, i}(\sigma_{m, i}, {\bf r}, t)$ induced by $\sigma_{m, i}$ at any position $\bf r$ is given \cite{Henry2016}
\begin{widetext}
\begin{equation} \label{AnySurfFld}
\begin{aligned}
{\bf h}_{d,i}(\sigma_{m,i},{\bf r},t) &= -\nabla \left [\psi_{m,i}(\sigma_{m,i},0,{\bf r},t)+\psi_{m,i}(\sigma_{m,i},-d,{\bf r},t)\right]\\&=\left\{
\begin{aligned}
 &\frac{m_{y,i}}{2}e^{-|k_x|y}\left (1-e^{-|k_x|d}\right )\left [j\sign {k_x} {\bf x}+ {\bf y}\right ],&y\geq0\\
 &-\frac{j m_{y,i}}{2}\left [e^{|k_x|y}-e^{-|k_x|(y+d)}\right ]\sign {k_x} {\bf x}+\frac{m_{y,i}}{2}\left [-e^{-|k_x|(y+d)}-e^{|k_x|y}\right ] {\bf y},&-d\leq y<0\\
  &\frac{m_{y,i}}{2}e^{|k_x|y}\left (e^{|k_x|d}-1\right )\left [-j\sign {k_x} {\bf x}+ {\bf y}\right ].&y<-d\\
\end{aligned}
\right.
\end{aligned}
\end{equation}
\end{widetext}
Finally, we obtain the dipolar magnetic field ${\bf h}_{d,i}({\bf r},t) = {\bf h}_{d,i}(\rho_{m,i},{\bf r},t)+{\bf h}_{d,i}(\sigma_{m,i},{\bf r},t)$ in the whole space
\begin{widetext}
\begin{equation} \label{AnyDipFld}
{\bf h}_{d,i}({\bf r},t) =\left\{
\begin{aligned}
&\frac{1}{2}e^{-|k_x|y}\left (1-e^{-|k_x|d}\right )\left\{\left [-m_{x,i}+j\sign {k_x}m_{y,i}\right ] {\bf x}+\left [j\sign {k_x}m_{x,i}+m_{y,i}\right ] {\bf y}\right\},&y\geq0\\
 & \left\{\left [e^{-|k_x|(y+d)}+e^{|k_x|y}-2\right ]\frac{m_{x,i}}{2}+\frac{j\sign {k_x}}{2}\left [e^{-|k_x(y+d)|}-e^{-|k_xy|}\right ]m_{y,i}\right \} {\bf x}&\\
 &+\left \{\frac{j\sign {k_x}}{2}\left [e^{|k_x|y}-e^{-|k_x|(y+d)}\right ]m_{x,i}-\left [e^{-|k_xy|}+e^{-|k_x(y+d)|}\right ]\frac{m_{y,i}}{2}\right\} {\bf y},&-d \leq y<0\\
  &\frac{1}{2}e^{|k_x|y}\left (e^{|k_x|d}-1\right )\left \{\left [-m_{x,i}-j\sign {k_x}m_{y,i}\right ] {\bf x}+\left [-j\sign {k_x}m_{x,i}+m_{y,i}\right ] {\bf y}\right \},&y<-d.\\
\end{aligned}
\right.
\end{equation}
\end{widetext}
\subsection{Self demagnetizing factors} \label{SelfDE}
When calculating the demagnetizing factor of a single layer with thickness $d_i$, we care about the region $-d_i<y<0$. The demagnetizing factors inside the film are defined as the ratios between the average dipolar field and the magnetization
\begin{subequations} \label{DemFacDef}
\begin{align}
\label{DemFacDefa}-n_{x,i}m_{x,i}=\frac{1}{d}\int_{-d}^{0} {\bf x}\cdot{\bf h}_{d,i}({\bf r},t)dy, \\
\label{DemFacDefb}-n_{yx}m_{x,i}=\frac{1}{d}\int_{-d}^{0} {\bf y}\cdot{\bf h}_{d,i}({\bf r},t)dy, \\
\label{DemFacDefc}-n_{xy}m_{x,i}=\frac{1}{d}\int_{-d}^{0} {\bf x}\cdot{\bf h}_{d,i}({\bf r},t)dy, \\
\label{DemFacDefd}-n_{y,i}m_{x,i}=\frac{1}{d}\int_{-d}^{0} {\bf y}\cdot{\bf h}_{d,i}({\bf r},t)dy. &
\end{align}
\end{subequations}
We obtain
\begin{subequations} \label{DemFac}
\begin{align}
\label{DemFaca}n_{x,i}=1-&n_{y,i}=1-\frac{1-e^{-|k_x|d_i}}{|k_x|d_i} \\
\label{DemFacb}n_{xy}&=n_{yx}=0.
\end{align}
\end{subequations}
The net self-induced dipolar field ${\bf h}_{d,self,i}=h_{d,self,x} {\bf x}+h_{d,self,y} {\bf y}$ by the SWs can be evaluated
\begin{equation} \label{DipFldSelf}
\left[
 \begin{aligned}
 h_{d,self,x} \\
 h_{d,self,y}
 \end{aligned}
\right]=
-\left[
\begin{matrix}
     n_{x,i} & 0 \\
     0 & n_{y,i}
\end{matrix}
\right]
\left[
 \begin{aligned}
 m_{x,i} \\
 m_{y,i}
 \end{aligned}
\right].
\end{equation}
The ratio $\varepsilon_{hd}$ between $h_{d,x}$ and $h_{d,y}$ is given as
\begin{equation} \label{EllipSelf}
\varepsilon_{hd}=\frac{h_{d,x}}{h_{d,y}}=\frac{n_{x,i}m_{x,i}}{n_{y,i}m_{y,i}},
\end{equation}
indicating that the chirality of the self-induced dipolar field depends on the chirality of the dynamic magnetization.
\subsection{Mutual demagnetizing factors}\label{MutDE}
\begin{figure}[htbp!]
  \centering
  \includegraphics[width=0.4\textwidth]{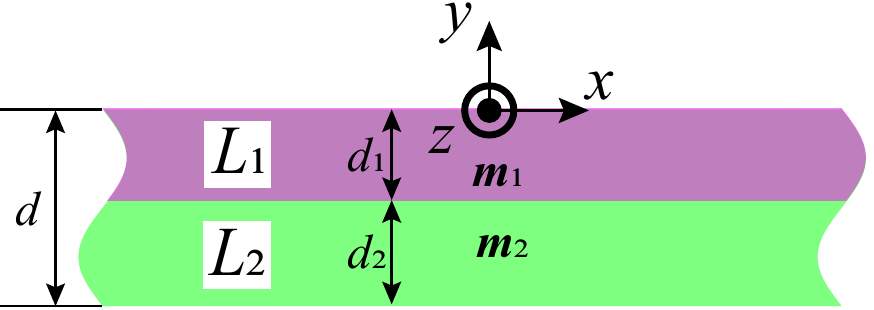}\\
  \caption{Schematic of the bilayer consisted of $L_1$ (purple) and $L_2$ (green) with SWs ${\bf m}_1$ and ${\bf m}_2$ inside, respectively.} \label{SMfig2}
\end{figure}
In this part, we consider the dipolar effects between the two adjacent layers labelled as $L_1$ and $L_2$, as shown in Fig. \ref{SMfig2}. They are located from $y = -d_1$ to 0 and from $y=-d$ to $-d_1$, respectively. For simplification, we denote $d_2 = d-d_1$. The SWs propagating inside take the form ${\bf m}_p = {\bf m}_{0,p} e^{j(\omega t-k_{x,p}x)} = m_{x,p} {\bf x}+m_{y,p} {\bf y}$ with $p = 1,2$. According to Eq. (\ref{AnyDipFld}), the dipolar field induced by ${\bf m}_1$ and acting on $L_2$$\left [-(d_1+d_2)< y < -d_1\right ]$ is given as
\begin{widetext}
\begin{equation} \label{DipFld12}
\begin{aligned}
{\bf h}_{d,12}({\bf r},t)
 =  \frac{1}{2}e^{|k_{x,1}|y}\left (e^{|k_{x,1}|d_1}-1\right )\left\{\left [-m_{x,1}-j\sign{k_{x,1}}m_{y,1}\right ] {\bf x}
 +\left [-j\sign{k_{x,1}}m_{x,1}+m_{y,1}\right ] {\bf y}\right \}.
\end{aligned}
\end{equation}
\end{widetext}
The average dipolar field acting on $L_2$ can be evaluated by introducing the mutual demagnetizing factors $n_{x12}$, $n_{xy12}$, $n_{yx12}$ and $n_{y12}$
\begin{subequations} \label{GenDemFacDef12}
\begin{align}
\label{GenDemFacDef12a}-n_{x12}m_{x,1}-n_{xy12}m_{y,1}=\frac{1}{d_2}\int_{-(d_1+d_2)}^{-d_1} {\bf x}\cdot{\bf h}_{d,12}({\bf r},t)dy, \\
\label{GenDemFacDef12b}-n_{yx12}m_{x,1}-n_{y12}m_{y,1}=\frac{1}{d_2}\int_{-(d_1+d_2)}^{-d_1} {\bf y}\cdot{\bf h}_{d,12}({\bf r},t)dy. &
\end{align}
\end{subequations}
We obtain
\begin{subequations} \label{GenDemFac12}
\begin{align}
\label{GenDemFac12a}&n_{x12}=-n_{y12}
=\frac{\left(1-e^{-|k_{x,1}|d_1}\right)\left(1-e^{-|k_{x,1}|d_2}\right)}{2|k_{x,1}|d_2}, \\
\label{GenDemFac12b} &n_{xy12}=n_{yx12}=j\sign{k_{x,1}}n_{x12}.
\end{align}
\end{subequations}
The dipolar field induced by ${\bf m}_2$  and acting on $L_1 (-d_1<y<0)$ is given as
\begin{widetext}
\begin{equation} \label{DipFld21}
{\bf h}_{d,21}({\bf r},t)
=\frac{1}{2}e^{-|k_{x,2}|(y+d_1)}\left (1-e^{-|k_{x,2}|d_2}\right )\left\{\left [-m_{x,2}+j\sign{k_{x,2}}m_{y,2}\right ]{\bf x}
+\left [j\sign{k_{x,2}}m_{x,1}+m_{y,2}\right ]{\bf y}\right \}.
\end{equation}
\end{widetext}
Similarly, we introduce $n_{x21}$, $n_{xy21}$, $n_{yx21}$ and $n_{y21}$
\begin{subequations} \label{GenDemFacDef21}
\begin{align}
\label{GenDemFacDef21a}-n_{x21}m_{x,1}-n_{xy21}m_{y,1}=\frac{1}{d_1}\int_{-d_1}^{0} {\bf x}\cdot{\bf h}_{d,21}({\bf r},t)dy, \\
\label{GenDemFacDef21b}-n_{yx21}m_{x,1}-n_{y21}m_{y,1}=\frac{1}{d_1}\int_{-d_1}^{0} {\bf y}\cdot{\bf h}_{d,21}({\bf r},t)dy. &
\end{align}
\end{subequations}
We obtain
\begin{subequations} \label{GenDemFac21}
\begin{align}
\label{GenDemFac21a}&n_{x21}=-n_{y21}
=\frac{\left(1-e^{-|k_{x,2}|d_1}\right)\left(1-e^{-|k_{x,2}|d_2}\right)}{2|k_{x,2}|d_1}, \\
\label{GenDemFac21b} &n_{xy21}=n_{yx21}=-j\sign{k_{x,2}}n_{x21}.
\end{align}
\end{subequations}

Finally, the mutual net dipolar field ${\bf h}_{d,12}=h_{d,x,12} {\bf x}+h_{d,y,12} {\bf y}$ and ${\bf h}_{d,21}=h_{d,x,21} {\bf x}+h_{d,y,21} {\bf y}$ can be evaluated
\begin{equation} \label{DipFldMut12}
\left[
 \begin{aligned}
 h_{d,x,12} \\
 h_{d,y,12}
 \end{aligned}
\right]=-n_{x12}
\left[
\begin{matrix}
     1 & j\sign{k_{x,1}} \\
     j\sign{k_{x,1}} & -1
\end{matrix}
\right]
\left[
 \begin{aligned}
 m_{x,1} \\
 m_{y,1}
 \end{aligned}
\right],
\end{equation}
and
\begin{equation} \label{DipFldMut21}
\left[
 \begin{aligned}
 h_{d,x,21} \\
 h_{d,y,21}
 \end{aligned}
\right]=-n_{x21}
\left[
\begin{matrix}
     1 & -j\sign{k_{x,2}} \\
     -j\sign{k_{x,2}} & -1
\end{matrix}
\right]
\left[
 \begin{aligned}
 m_{x,2} \\
 m_{y,2}
 \end{aligned}
\right].
\end{equation}
The ratio $\varepsilon_{hd12}$ ($\varepsilon_{hd21}$) between $h_{d,x,12}$ and $h_{d,y,12}$ ($h_{d,x,21}$ and $h_{d,21,y}$) is given as
\begin{subequations} \label{EllipMut}
\begin{align}
\label{EllipMuta}\varepsilon_{hd12}=\frac{m_{x,1}+j\sign{k_{x,1}}m_{y,1}}{j\sign {k_{x,1}}m_{x,1}-m_{y,1}}=-j\sign{k_{x,1}}, \\
\label{EllipMutb} \varepsilon_{hd21}=\frac{m_{x,2}-j\sign{k_{x,2}}m_{y,2}}{-j\sign {k_{x,2}}m_{x,2}+m_{y,2}}=j\sign{k_{x,2}}.
\end{align}
\end{subequations}
indicating that the chirality of the mutual dipolar field depends on the signs of the wave vectors. The net mutual demagnetizing field can be estimated as
\begin{widetext}
\begin{equation} \label{DeMut}
\begin{aligned}
{\bf h}_{d,mut}&=\frac{{\bf h}_{d,12}d_2+{\bf h}_{d,21}d_1}{d}\\
&=-\frac{\left(1-e^{-|k_{x,2}|d_1}\right)\left(1-e^{-|k_{x,2}|d_2}\right)}{2|k_x|d}\left\{
 \begin{aligned}
 \left[(m_{x,1}+m_{x,2})+j\sign {k_x}(m_{y,1}-m_{y,2})\right] {\bf x}  +\left[j\sign {k_x}(m_{x,1}-m_{x,2})-(m_{y,1}+m_{y,2})\right] {\bf y}
 \end{aligned}
\right\}.
\end{aligned}
\end{equation}
\end{widetext}

Here, we note that in the main text, the dynamic magnetization ${\bf m}_1$ and ${\bf m}_2$ satisfy the boundary condition ${\bf m}_1={\bf m}_2\bigg |_{y=-d_1}$ \cite{Verba2020}, which gives $k_{x,1} = k_{x,2} = k_{x}$.

\end{appendix}

\end{document}